\documentclass[letter paper, 10 pt, conference]{ieeeconf}
\IEEEoverridecommandlockouts                              

\usepackage[utf8]{inputenc}
\usepackage{cite}
\usepackage[utf8]{inputenc}
\usepackage{cite}
\usepackage{qtree}
\usepackage{tikz}
\usepackage{amsmath,amssymb,amsfonts}
\usepackage{eufrak}
\usepackage{dsfont}
\usepackage{steinmetz}
\usepackage{algorithmicx}
\usepackage{algorithm}
\usepackage{algpseudocode}
\usepackage{graphicx}
\usepackage{amsmath,amssymb}
\usepackage{textcomp}
\usepackage{algpseudocode} 
\usepackage{mathtools}
\usepackage{nicefrac}
\usepackage{txfonts}
\usepackage{tikz}
\usepackage{dsfont}
\usepackage{graphicx}
\usepackage{amsmath,amssymb}
\usepackage{xcolor}
\usepackage{mathtools}
\usepackage{nicefrac}
\usepackage{txfonts}
\usepackage{todonotes}
\usepackage{cleveref}
\usepackage{titlesec}
\usepackage{subfig}
\usepackage[font=small]{caption}

\newcommand{\R}{\mathbb{R}}

\newcommand{\tr}{{\rm{tr}}}

\newcommand{\Q}{\mathcal{Q}}

\titlespacing*{\subsection}
  {0pt}{0.4\baselineskip}{0.2\baselineskip}

 \titlespacing*{\section}
   {0pt}{0.7\baselineskip}{0.3\baselineskip}

\renewcommand{\baselinestretch}{1}
\allowdisplaybreaks

\DeclareMathOperator{\argmin}{\arg\!\min}

\def\mb{\mathbb}
\def\mc{\mathcal}
\def\beq{\begin{equation*}}
\def\eeq{\end{equation*}}
\def\bql{\begin{equation}}
\def\eql{\end{equation}}
\def\bqn{\begin{eqnarray*}}
\def\eqn{\end{eqnarray*}}
\def\bnl{\begin{eqnarray}}
\def\enl{\end{eqnarray}}
\def\bna{\bql\begin{array}{rcl}}
\def\ena{\end{array}\eql}
\def\bnn{\beq\begin{array}{rcl}}
\def\enn{\end{array}\eeq}
\def\bma{\begin{bmatrix}}
\def\ema{\end{bmatrix}}
\def\bmx{\begin{matrix}}
\def\emx{\end{matrix}}
\def\ben{\begin{enumerate}}
\def\een{\end{enumerate}}
\def\bit{\begin{itemize}}
\def\eit{\end{itemize}}
\def\bei{\begin{itemize}}
\def\eei{\end{itemize}}
\def\bet{\begin{tabular}}
\def\eet{\end{tabular}}

\newtheorem{thm}{Theorem}

\newtheorem{df}{ Definition}

\newtheorem{rem}{ Remark}

\newtheorem{asm}{Assumption}

\title{\Large \bf Data to Certificate: Guaranteed Cost Control with Quantization-Aware System Identification}

\author{Shahab Ataei, Dipankar Maity, and Debdipta Goswami 
\thanks{S. Ataei is with the Department of Electrical and Computer Engineering, The Ohio State University, Columbus,
OH, 43210, USA (e-mail:{ \tt ataei.3@osu.edu}).}
\thanks{D. Maity is with the Department of Electrical and Computer Engineering and an affiliated faculty of the North Carolina Battery Complexity, Autonomous Vehicle, and Electrification Research Center (BATT CAVE), University of North Carolina at Charlotte,  NC, 28223, USA (e-mail: {\tt {dmaity@charlotte.edu}}).
}
\thanks{D. Goswami is with the Department of Mechanical and Aerospace Engineering, The Ohio State University, Columbus,
OH, 43210, USA (e-mail:{ \tt goswami.78@osu.edu}).
}
}

\begin{document}
\maketitle
\thispagestyle{empty}
\pagestyle{empty}

\begin{abstract}
Cloud-assisted system identification and control have emerged as practical solutions for low-power, resource-constrained control systems such as micro-UAVs. In a typical cloud-assisted setting, state and input data are transmitted from local agents to a central computer over low-bandwidth wireless links, leading to quantization. 
This paper investigates the impact of state and input data quantization on a linear time invariant (LTI) system identification, derives a worst-case bound on the identification error, and develops a robust controller for guaranteed cost control. We establish a fundamental bound on the model error that depends \emph{only} on the quantized data and quantization resolution, and develop a linear matrix inequality (LMI) based guaranteed cost robust controller under this error bound. 

\end{abstract}


\section{Introduction} 
The evolution of autonomous and multi-agent systems toward safety-critical applications highlights the need for control strategies that ensure reliable decision-making under uncertainty \cite{zarei2025MAS}. The inherent complexity and uncertainty arising from the rich dynamics of such systems introduce levels of complexity that traditional modeling cannot efficiently represent or exploit for robust performance \cite{Vaziri2025CNN}. Consequently, there is a strong need for modern, data-driven system identification and control methods that can learn from operational data while supporting verifiable performance and safety guarantees \cite{hedesh2024NN,alavi2025safety}. Nevertheless, the sophistication that enables such capabilities often comes at a high computational cost. Existing studies often assume ample onboard computational power, an assumption that fails on resource-constrained platforms such as low-power, lightweight robots \cite{Cleary2020}.

One common approach to address these computational limitations involves offloading computationally intensive control algorithms to cloud or edge computing platforms that possess sufficient processing capacity \cite{Li2023cloud}. A widely adopted alternative is to simplify the control task by employing onboard linear controllers with little computation. A linear controller can be designed either by linearizing the system dynamics \cite{Schweickhardt2009Linearization} that requires system knowledge or by leveraging data-driven methods to estimate linear system approximations, such as the Koopman operator-based methods~\cite{korda2018}. Such linear controllers can be synthesized using simple Linear Matrix Inequality (LMI) design methods that accommodate a range of scenarios, as demonstrated in \cite{hedesh2025LMI, Ganji2024LMI}, and are computationally efficient, making them well suited to resource-constrained hardware. However, the data-driven system identification phase that precedes controller design remains computationally demanding and is typically executed remotely on more capable computing platforms.
This cloud-assisted identification paradigm requires transmitting input/state data from local agents to a central processor over resource-limited communication networks, which induce time delays, packet disordering, and—critically—quantization \cite{Zhang2013NCSConstraints}. Because identification typically operates off-line or without hard real-time deadlines, quantization becomes the dominant bottleneck degrading data quality. At the same time, judicious quantization is essential for meeting bandwidth/energy budgets and curbing downstream computation in resource-limited, lightweight robotic applications~\cite{Danaee2018QuantizationEnergy}.

%

%

For a discrete time linear time-invariant (LTI) system, system matrices are typically identified via least-squares optimization using input and state data snapshots \cite{ljung1998system}. While the accuracy of least-squares estimates improves with data volume \cite{Arbabi2017, Ziemann2023} and worsens with noise \cite{Ziemann2023}, the impact of bandwidth limitations—i.e., quantization—on estimation remains unclear. The downstream effect of identification errors from degraded data on controller performance is also largely unexamined. Since identification often precedes control in unknown systems, quantization errors can cascade through estimators and controllers, affecting overall performance. The choice of quantizer is therefore critical \cite{maity2021optimal, maity2023optimal}.

In this paper, we study the effects of quantization on the data-driven system identification and robust control of unknown LTI systems. Our prior works \cite{maity2024effect, maity2024EDMD, ataei2025koopman} delved into the effect of dither quantization on Koopman-based system identification method. Here, we derive an explicit norm bound on the identification error \emph{solely from the quantized data and quantization resolution} with no prior knowledge on unquantized data and true system matrices. This contrasts with our previous work \cite{ataei2025qsid} that derived an identification error bound as a function of unquantized data and true system matrices.
%
%
Finally, we utilize LMIs to design a guaranteed cost robust controller with a finite bound on infinite horizon quadratic cost for all possible identification error within the derived norm bound.
To the best of our knowledge, this is the first attempt to quantify the effect of quantization on system identification solely from quantized data and develop a robust controller to mitigate the uncertainty.

The main contributions of this paper are as follows: (i) We establish explicit norm bounds on the identification error arising from quantized input and state data in the identification of LTI systems. In particular, we derive a norm bound on identification error \emph{solely from the quantized data and quantization resolution} with no prior knowledge whatsoever about the true data and system matrices. It is also evident that the error bound scales down logarithmically with the quantization word-length $b$. (ii) We develop a guaranteed cost robust control for the identified system with a norm bounded uncertain identification error using an LMI-based convex programming.

The rest of the paper is organized as follows: We define our problem statement in \Cref{sec:ProblemStatement} and analyze the effect~of quantization on system identification in~\Cref{sec:QuantizedDMD}, deriving the norm bound of the~identification error. 
Then we proceed to derive a guaranteed cost robust controller for the identified system in \Cref{sec:MPC}.
We discuss our observations from implementing the robust controller on two dynamical systems in \Cref{sec:Validation} and we conclude the paper in \Cref{sec:conclusions}.\\

\textit{Notations:} Set of non-negative integers are denoted by $\mb{N}_0$. 
$(\cdot)^{\dagger}$ and $(\cdot)^\top$ denote the Moore--Penrose inverse and transpose of a matrix, respectively. For a matrix $M$, $\sigma_{\min}(M)$ and $\sigma_{\max}(M)$ represent its minimum and maximum singular values. If $M$ is symmetric, $\lambda_{\min}(M)$ and $\lambda_{\max}(M)$ denote its minimum and maximum eigenvalues.
We use $x^\top M x=x^\top M(\bullet)$ for brevity.
$\|\cdot\|$ denotes a norm, where we use Euclidean norm for vectors and Frobenius norms for matrices. 
The Big-O notation is denoted by $O(\cdot)$.

\begin{figure}[t]
\centering 
\includegraphics[trim=0cm 0cm 0cm 0cm, clip=true, width=0.45\textwidth]{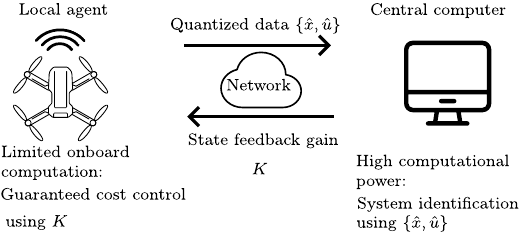}
\caption{Framework Overview.} \label{Fig:framework}
\end{figure}

\section{Problem Statement} \label{sec:ProblemStatement}

\Cref{Fig:framework} outlines the architecture considered in this work. During the system identification phase, each resource‑constrained agent collects state–input snapshots, applies quantization, and transmits the quantized data to a central computer over a bandwidth‑limited network. The central computer then runs a least-squares identification algorithm to estimate system matrices $\hat{A},\hat{B}$.
In particular, we consider a discrete-time LTI system \vspace{-5pt}
\begin{align} \label{eq:dynamics}
\begin{split}
    x_{t+1} = Ax_t + Bu_t,\ x_t\in\mc{M}\subset \mb{R}^n,\ u_t\in\mc{U}\subset\mb{R}^m,\  t\in\mb{N}_0,
\end{split}
\end{align}
where both $A$ and $B$ matrices are unknown. 
Using the identified model \((\hat A,\hat B)\), we design a state-feedback law \(u_t=-Kx_t\) that guarantees an upper bound on the infinite-horizon quadratic cost
\begin{align} \label{eq:cost}
    J\left(x_0\right)=\lim _{n \rightarrow \infty} \sum_{t=0}^n x_t^\top Q x_t+u_t^\top R u_t \nonumber\\
    \text{subject to: } x_{t+1} = Ax_t + Bu_t
\end{align}
where $Q \in \mb{R}^{n \times n}, R \in \mb{R}^{m \times m}, Q \succ 0, R \succ 0$, and
\begin{align*}
\left[\begin{array}{cc}
Q & 0 \\
0 & R
\end{array}\right]=\left[\begin{array}{c}
C_c^\top \\
D_c^{u T}
\end{array}\right]\left[\begin{array}{ll}
C_c & D_c^u
\end{array}\right] \succeq 0
\end{align*}
and $C_c \in \mb{R}^{(n+m) \times n}$ and $D_c^u \in \mb{R}^{(n+m) \times m}$ denote the factorization of the cost function.
In this problem, we investigate the scenario where the system identification of $({A}, {B})$ is performed with \textit{quantized data} and a least-squares algorithm akin to the Dynamic Mode Decomposition method \cite{schmid2010}. 
While the least-squares problem---to be described soon---is able to \textit{accurately} identify the system under unquantized data, it fails to attain such perfect accuracy under quantized data. 
First objective of this work is to investigate the effect of the quantization on the system identification and derive an explicit norm bound on the identification error \emph{solely from the quantized data and quantization resolution} with no prior knowledge of the true data and system matrices. This contrasts with our previous work \cite{ataei2025qsid} that derived identification error bounds as a function of true system matrices and unquantized data (cf. \cite{ataei2025qsid}, Theorem 1). Furthermore, we wish to design a guaranteed cost robust controller for the true LTI system using the identified system matrices for all possible identification error within the derived bound.
Before proceeding further, let us assume the following on the original system \eqref{eq:dynamics}.

\begin{asm}\label{assm:controllability}
    $(A,B)$ is a controllable pair.
\end{asm}
This assumption is instrumental, as regulation may otherwise be unachievable, even with a perfectly identified system. 

\begin{rem}
    The effect of quantization on controller synthesis is a well studied problem \cite{fu2024tutorial}. However, quantization effects on system identification and it's effect on a robust controller design is not yet explored.
    In this paper, we will provide a guaranteed cost stabilizing controller that is robust with respect to the identification error due to quantization.
\end{rem}



\subsection{System Identification}

The system matrices $A$ and $B$ are identified from \textit{quantized} input and state data snapshots $\{\tilde{x}_t\}_{t=0}^T$ and $\{\tilde{u}_t\}_{t=0}^{T-1}$ as follows: \vspace{-10pt}
\begin{align}\label{Eq: optimization}
\begin{split}
    \hat G = [\hat A,\,\hat B]=& \underset{\mc{A} {\in \R^{n \times n}},\,\mc{B} \in {\R^{n \times m}}}{\argmin} \tfrac{1}{T}\|\tilde X^{+} - \mc{A} \tilde X - \mc{B} \tilde U\|^2 \\
    = & \underset{\mc{G} \in {\R^{n \times (n+m)}}}{\argmin} \tfrac{1}{T}\|\tilde X^{+} - \mc{G} \tilde \Psi\|^2,
    \end{split}
\end{align}\vspace{-15pt}
\begin{align}
\label{eq:dataMatrix2}
\begin{split}
    \text{where }{\tilde X} = [\tilde x_0 ~ \hdots ~ \tilde x_{T-1}],\ &
    \tilde X^{+} = [\tilde x_1 ~ \hdots ~ \tilde x_{T}]\\
    {\tilde U} = [\tilde u_0~\hdots~\tilde u_{T-1}], & \text{ and }
    \tilde \Psi  = \begin{bmatrix}
        \tilde X^\top, 
        \tilde U^\top
    \end{bmatrix}^\top,
\end{split}
\end{align}
with $\tilde x_t$ and $\tilde u_t$ denoting the quantized versions of $x_t$ and $u_t$, respectively. 
Exact relationships between the unquantized and quantized variables will be provided later in~\Cref{sec:QuantizedDMD}. 

The control input $\{u_t\}_{t=0}^{T-1}$ can either be open-loop or closed-loop. However, system identification even without quantization needs some assumption on the data,\hspace{-1pt} namely the \textit{persistent excitation} criteria. 
Let $\Psi_{\rm uqz}$ denote the equivalent of $\tilde\Psi$ in \eqref{eq:dataMatrix2}, which consists of the unquantized data.  
\begin{asm}
    $\Psi_{\rm uqz}\Psi_{\rm uqz}^\top \succ 0$.
\end{asm}

\begin{rem} \label{rem:least-square}
    When data is not quantized and an adequate amount of data is present, the least-squares problem \eqref{Eq: optimization} has a unique solution $\mathcal{G}^* = [A, B]$. 
    For instance assume~$u_t = 0$ for all $t$, then the least-squares problem boils down~to\\ $\min_{\mathcal{A}} \tfrac{1}{T}\|X^+ - \mathcal{A}X\|^2 = \min_{\mathcal{A}}  \tfrac{1}{T} \sum_{t=0}^{T-1}\|Ax_t - \mathcal{A}x_t\|^2 = \min_{\mathcal{A}}  \tr\left((A-\mathcal{A})\left(\tfrac{1}{T}\sum_{t=0}^{T-1}x_t x_t^\top\right) (A-\mathcal{A})^\top \right)$.\\ 
    If enough data is accumulated to ensure $\tfrac{1}{T}\sum_{t=0}^{T-1}x_t x_t^\top \succ 0$, then $\mathcal{A}^* = A$ is the \textit{unique} optimal solution.
    Once matrix $A$ is identified, we may identify matrix $B$ in a similar fashion with enough data to ensure $\tfrac{1}{T}\sum_{t=0}^{T-1}u_t u_t^\top \succ 0$.
\end{rem}

\section{Effects of Quantization on $(\hat{A}, \hat{B})$} \label{sec:QuantizedDMD}

In this work, we adopt a \emph{memoryless scalar quantization (MSQ)} scheme. Our motivation is twofold: (i) We seek a worst-case model of the quantization effect so that the identification error can be captured by a deterministic bound \([\Delta A\ \ \Delta B]=:\Delta G\); (ii) This bound is the key ingredient needed to certify robustness of the subsequent controller. In MSQ, the quantizer is a static, deterministic mapping \(Q(\cdot)\) applied componentwise to the state–input snapshots.

More precisely, let $\Q$ be a quantizer with range $[s_{\min}, s_{\max}] \subseteq \R$ and resolution $\epsilon_q$ such that for any $s \in \R$ \vspace{-5pt}
\begin{align*}
    \Q(s) = \begin{cases} s_{\min}+
        \epsilon_q \left\lfloor\tfrac{s-s_{\min}}{\epsilon_q} \right\rfloor, \qquad & s \in [s_{\min}, s_{\max}], \\
        s_{\min}, & s< s_{\min}, \\
        s_{\min}  +
        \epsilon_q \left\lfloor\tfrac{s_{\max}-s_{\min}}{\epsilon} \right\rfloor & s> s_{\max}.
    \end{cases}
\end{align*}
Quantizer $\Q$ requires $b = \lceil\log_2\tfrac{s_{\max}-s_{\min}}{\epsilon_q}\rceil$ bits to represent its quantized output.  
Under an MSQ scheme, the quantized version of $s$ is denoted as 
$\tilde s = \Q(s)$. Consequently, the quantization error is defined as $e(s) \triangleq s-\tilde s = s - \Q(s)$. 
\begin{df}
Let $\tilde{x} \in \R^n$ and $\tilde{u} \in \R^m$ be the quantized representations of the state $x$ and input $u$, respectively. We define the vector of maximum component-wise quantization error bounds for the state as $e_x \in \R^n$, where its $i$-th element is $e_{x,i} \triangleq \sup_{t\in\{0,\dots,T\}} |\,\tilde x_{t,i} - x_{t,i}\,|$. Similarly, the vector of maximum component-wise quantization error bounds for the input is $e_u \in \R^m$, with its $j$-th element defined as $e_{u,j} \triangleq \sup_{t\in\{0,\dots,T-1\}} |\,\tilde u_{t,j} - u_{t,j}\,|$. We then define the aggregate error bounds $\varepsilon_x \triangleq \|e_x\|$ and $\varepsilon_u \triangleq \|e_u\|$. Finally, we let $\varepsilon \triangleq \sqrt{\varepsilon_x^2+\varepsilon_u^2}$ denote the maximum possible quantization error for $[x^\top ,\ u^\top ]^\top $.
\end{df}
\begin{rem}
    The value of each $e_{x,i}$ and $u_{x,j}$ is bounded by:
    \begin{align*}
        e_{x,i} \leq \frac{1}{2}\frac{x_{i,max} - x_{i,min}}{2^b},\; e_{u,j} \leq \frac{1}{2}\frac{u_{j,max} - u_{j,min}}{2^b},
    \end{align*} 
    where $x_{i,max},\;u_{j,max} $ and $x_{i,min},\;u_{j,min}$ are the component wise maximums and minimums of $x$ and $u$, respectively.
    Consequently, the logarithm of quantization error decreases, with a slope of $-\log 2 = -0.301$, as the number of bits increases.
\end{rem}

\begin{df}
    We let $E_{X^+}$ and $E_\Psi$ be the quantization error for data matrices defined in \eqref{Eq: optimization}, respectively. Therefore:
    \begin{align*}
        E_{X^+} = X_{\rm uqz}^+ -\tilde X^+,\;E_\Psi = \Psi_{\rm uqz} - \tilde \Psi.
    \end{align*}
\end{df}
To ensure that quantization saturation does not occur, we make the following assumption on the data.
\begin{asm} \label{asm:bounded}
   The data $\{x_t, u_t\}_{t=0}^T$ used in quantization is bounded.
   That is, there exists two scalars $q_{\min} $ and $ q_{\max}$  such that $q_{\min} \mathds{1} \preceq x_t \preceq q_{\max} \mathds{1}$ and $q_{\min} \mathds{1} \preceq u_t \preceq q_{\max} \mathds{1}$.
\end{asm}
Notice that, since the triplet $\{x_{t+1}, x_t, u_t\}$ represents a one time evolution of a linear system, the triplet is guaranteed to be bounded if the input and the start state are bounded. Therefore, \Cref{asm:bounded} holds true trivially. 
Furthermore, by scaling (down) the original data, one may reduce the magnitude of $q_{\min}$ and $q_{\max}$. 
That is, the scaled data  $\{cx_t, cu_t\}_{t=0}^T$, for $1>c>0$,  is bounded by $cq_{\min}$ and $cq_{\max}$. 
By choosing an appropriate $c$, one may ensure that $x_{\min} \le c q_{\min} < c q_{\max} \le x_{\max}$.
Consequently, no saturation occurs during quantization. 
Notice that, due to the linearity of the system dynamics, the scaled down data also follow the same dynamics \eqref{eq:dynamics} as the original data.

The main result of this section is summarized in the following theorem.

\begin{thm}[Norm bounded uncertainty]\label{thm:bounded_uncertainty}
If the following condition holds:
\begin{align} \label{eq:PEcondition}
    \sigma_{\min}(\tilde\Psi+E_\Psi) \ge \sigma_{\min}(\tilde\Psi)-\left(\sqrt{T}\right)\varepsilon,
\end{align}
then we can bound the system uncertainty matrices $\Delta G = [\Delta A, \Delta B]$ as follows:
\begin{align} \label{eq:G_bound}
    \|\Delta G\| \leq \rho,
\end{align}
where
\begin{align*}
    \rho &= \frac{T\left(\| \hat G\|_2\Gamma_W + \Gamma_Y\right)}{(\sigma_{\min}(\tilde\Psi)-(\sqrt{T})\varepsilon)^2},\\
    \Gamma_Y&=\frac{\sigma_{\max}(\tilde X^+)\varepsilon}{\sqrt{T}}+ 
    \frac{\sigma_{\max}(\tilde \Psi)\varepsilon_x}{\sqrt{T}}+\varepsilon_x\varepsilon, \text{ and }\\
    \Gamma_W&= \frac{2\varepsilon\sigma_{\max}(\tilde\Psi)}{\sqrt{T}}+\varepsilon^2.
\end{align*}

\end{thm}

\begin{proof}
The closed-form solution to the least-squares problem in \eqref{Eq: optimization} with quantized data is 
\begin{align} \label{eq:DMD_quantized}
   \hat G =  [\hat A, \hat B] =\frac{\tilde X^{+}  \tilde\Psi^\top}{T} \bigg(  \frac{\tilde\Psi  \tilde\Psi^\top}{T}\bigg)^{-1}, 
\end{align}
Similarly, for the unquantized data, the corresponding estimator is
\begin{align} \label{eq:DMD_unquantized}
    G =  [ A, B] =  \frac{X_{\rm uqz}^{+} \Psi_{\rm uqz}^\top}{T} \bigg( \frac{\Psi_{\rm uqz}  \Psi_{\rm uqz}^\top}{T} \bigg)^{-1}, 
\end{align}
where $[A, B]$ is the identified system (i.e., solution to \eqref{Eq: optimization}) with \textit{unquantized} data.
To quantify the uncertainty, we express the true system matrix $G$ in terms of the quantized data and the corresponding quantization error matrices:
\begin{align*}
    G =& ( \hat G + \Delta G) \\
    =& \frac{1}{T}(\tilde X^+ + E_{X^+})(\tilde \Psi+E_\Psi)^\top \bigg[\frac{1}{T}(\tilde \Psi+E_\Psi)(\tilde \Psi+E_\Psi)^\top\bigg]^{-1} \\
    =& \tfrac1T\big(\tilde X^+\tilde \Psi^\top + \tilde X^+E_\Psi^\top + E_{X^+}\tilde \Psi^\top + E_{X^+}E_\Psi^\top\big) \times \\
    &~~~~\big[\tfrac1T\big(\tilde \Psi\tilde \Psi^\top + \tilde \Psi E_\Psi^\top + E_\Psi\tilde \Psi^\top + E_\Psi E_\Psi^\top\big)\big]^{-1}.
\end{align*}
For notational simplicity, let us define the following terms:
\begin{align*}
    Y &= \tfrac1T\big(\tilde X^+\tilde \Psi^\top),\\
    \Delta Y & = \tfrac1T\big( \tilde X^+E_\Psi^\top + E_{X^+}\tilde \Psi^\top + E_{X^+}E_\Psi^\top\big),\\
    W &= \tfrac1T\tilde \Psi\tilde \Psi^\top,\\
    \Delta W &= \tfrac1T(\tilde \Psi E_\Psi^\top + E_\Psi\tilde \Psi^\top + E_\Psi E_\Psi^\top).
\end{align*}
Substituting these definitions yields the compact expression:
\begin{align*}
   G = Y(W+\Delta W)^{-1} + \Delta Y(W+\Delta W)^{-1}.
\end{align*}
To handle the matrix inverse, we apply the Woodbury matrix identity: $(A+B)^{-1} = A^{-1} - A^{-1}B(A+B)^{-1}$.
This identity allows us to separate the nominal terms from the error terms:
\begin{align*}
    G &= Y(W^{-1} - W^{-1}
\Delta W(W+\Delta W)^{-1})+\Delta Y(W+\Delta W)^{-1}\\
&=\tilde X^+\tilde \Psi^\top(\tilde \Psi\tilde \Psi^\top)^{-1} -\tilde X^+\tilde \Psi^\top (\tilde \Psi\tilde \Psi^\top)^{-1}\Delta W (\tilde \Psi\tilde \Psi^\top + \Delta W)^{-1}\\
&+ \Delta Y(\tilde \Psi\tilde \Psi^\top + \Delta W)^{-1}.
\end{align*}
Recall that $\hat{G} = \tilde X^+\tilde \Psi^\top(\tilde \Psi\tilde \Psi^\top)^{-1}$, and consequently, we have 
\begin{align*}
    \Delta G &= -\tilde X^+\tilde \Psi^\top (\tilde \Psi\tilde \Psi^\top)^{-1}\Delta W (\tilde \Psi\tilde \Psi^\top + \Delta W)^{-1} \\
    &+ \Delta Y(\tilde \Psi\tilde \Psi^\top + \Delta W)^{-1}.
\end{align*}
Next, we establish bounds on the norms of the error components $\Delta Y$ and $\Delta W$ using the triangle inequality and properties of matrix norms:
\begin{align*}
    \|\Delta Y\| &\leq \tfrac1T\big( \|\tilde X^+\|\|E_\Psi^\top\| + \|E_{X^+}\|\|\tilde \Psi^\top\| + \|E_{X^+}\|\|E_\Psi^\top\big\|)\\
    &\leq \frac{\sigma_{\max}(\tilde X^+)\sqrt{T}\varepsilon}{T}+ 
    \frac{\sigma_{\max}(\tilde \Psi)\sqrt{T}\varepsilon_x}{T}+\frac{\sqrt{T}\varepsilon_x \sqrt{T}\varepsilon}{T}=\Gamma_Y,\\
\end{align*}
\begin{align*}
    \|\Delta W\| &\leq\tfrac1T(\|\tilde \Psi\|\| E_\Psi^\top\| + \|E_\Psi\| \|\tilde \Psi^\top\| + \|E_\Psi\| \|E_\Psi^\top\|\\
    &\leq \frac{2\sigma_{\max}(\tilde\Psi)\sqrt{T}\varepsilon +T\varepsilon^2 } {T}=\Gamma_W.
\end{align*}
Using the variational characterization of singular values, we have:
\begin{align*}
    \sigma_{\min}(\tilde\Psi+E_\Psi) \geq \sigma_{\min}(\tilde\Psi)-\sigma_{\max}(E_\Psi).
\end{align*}
We also have $\sigma^2_{\max}(E_\Psi) = \lambda_{\max}(E_\Psi E_\Psi^\top) \leq T\varepsilon^2$. Therefore:
\begin{align*}
    \sigma_{\min}(\tilde\Psi+E_\Psi) \ge \sigma_{\min}(\tilde\Psi)-\left(\sqrt{T}\right)\varepsilon.
\end{align*}
Since $(\tilde \Psi+E_\Psi)(\tilde \Psi+E_\Psi)^\top$ is a positive definite matrix , we have :
\begin{align*}
     \left\|\left( \frac{(\tilde \Psi+E_\Psi)(\tilde \Psi+E_\Psi)^\top}{T}\right)^{-1}\right\| &=\frac{T} {\lambda_{\min} (\tilde \Psi+E_\Psi)(\tilde \Psi+E_\Psi)^\top}\\ 
     &=  \frac{T}{\sigma^2_{\min}(\tilde\Psi+E_\Psi)}\\
     &\leq \frac{T}{(\sigma_{\min}(\tilde\Psi)-(\sqrt{T})\varepsilon)^2}.
\end{align*}
Combining the bounds on $\|\Delta W\|$, $\|\Delta Y\|$, and and the inverse term yields:
\begin{align*}
    \|\Delta G\| \leq \frac{T\left(\| \hat G\|_2\Gamma_W + \Gamma_Y\right)}{\big(\sigma_{\min}(\tilde\Psi)-\varepsilon\sqrt{T}\big)^2} = \rho.
\end{align*}

\end{proof}
\begin{rem}[Robust Persistence of Excitation]
The condition in \eqref{eq:PEcondition} can be interpreted as a robust version of the classical Persistence of Excitation (PE) condition required for system identification. The term $\sigma_{\min}(\tilde\Psi)$ quantifies the information content of the data along its least-excited direction, whereas the standard PE condition simply requires this value to be strictly positive ($\sigma_{\min}(\Psi_{\rm uqz}) > 0$). In contrast, the term \textbf{$\sqrt{T}\varepsilon$} represents the aggregated uncertainty or worst-case ``noise'' magnitude introduced by the quantization process. Therefore, the condition essentially requires that the signal strength in the weakest direction of the data must exceed the maximum noise level introduced by quantization. This ensures the identification problem remains well-conditioned, demanding that the data be sufficiently rich to overcome the specific level of quantization noise.
\end{rem}

\Cref{thm:bounded_uncertainty} shows how the quantization resolution affects the identification process, providing a fundamental connection between data quality and identification accuracy.





\begin{rem}
   The identification error diminishes as $\varepsilon \to 0$. 
    Furthermore, since the quantization resolution $\varepsilon$ is proportional to $2^{-b}$, where $b$ is the word-length of the quantizer, the system identification errors decay exponentially with the word-length---a phenomenon that will be evident in all simulation results. 
\end{rem}


\section{Guaranteed Cost Controller under Identification Error}\label{sec:MPC}

This section establishes quadratic (robust) stability and a performance certificate for the LTI system \eqref{eq:dynamics} rendered uncertain by the identification error $[\Delta A,~\Delta B]$ due to quantization. Substituting the norm bounded error matrices $[\Delta A,\Delta B]$ obtained in \Cref{thm:bounded_uncertainty} in \eqref{eq:dynamics} yields:
\begin{align} \label{eq:dyanmics_uncertainty}
    x_{t+1} = (\hat A+\Delta A)x_t + (\hat B+\Delta B)u_t.
\end{align}
\begin{rem}
    The system in \eqref{eq:dyanmics_uncertainty} is an uncertain LTI system pertaining to the fact that $[\Delta A, \Delta B]$ is unknown. Only a norm bound $\rho$ can be obtained for the identification error matrices as demonstrated in \Cref{thm:bounded_uncertainty}.
\end{rem}
Here, we want to develop a robust controller for \eqref{eq:dyanmics_uncertainty} that stabilizes the system with a bounded infinite-horizon cost for all possible $[\Delta A, \Delta B]$ such that $\left\|[\Delta A, \Delta B]\right\|\leq \rho$.
This leads to the formal definition of a Guaranteed Cost Controller for the uncertain system.
\begin{df}\label{df:GCC}
A state feedback controller $u_t = -Kx_t$ is said to be a \emph{stabilizing Guaranteed Cost Controller} for the uncertain system \eqref{eq:dyanmics_uncertainty} if there exists a symmetric matrix $P \succ 0$ that upper-bounds the cost functional \eqref{eq:cost} for 
\begin{align*}
(\hat A+\Delta A-(\hat B+\Delta B) K)^\top P (\bullet) - P + Q + K^\top R K \preceq\ 0,
\end{align*}
such that $J\left(x_0\right) \leq J^*\left(x_0\right)\triangleq x_0^\top P x_0$ for the closed loop system
\begin{align} \label{eq:closedloop}
    x_{t+1} = (\hat{A} + \Delta A)x_t - (\hat{B} + \Delta B)Kx_t
\end{align}
and all admissible uncertainties satisfying $\|\Delta G\| \le \rho$. The value $x_0^\top P x_0$ is referred to as the \textit{guaranteed cost}.
\end{df}
\begin{rem}
    \Cref{assm:controllability} ensures that $[\hat A+\Delta A,\,\hat B+\Delta B]$ form a controllable pair. 
\end{rem}
We now proceed to obtain a stabilizing guaranteed cost controller. 
The uncertain system \eqref{eq:dyanmics_uncertainty} can be equivalently written as
\begin{align}\label{eq:dyanmics_uncertainty_equivalent}
    x_{t+1} &= \hat Ax_t + \hat Bu_t + B^ww_t, \nonumber\\
    z_k &= C_zx_t +D_z^uu_t, \nonumber \\
    \text{subject to: } w_t &= \Delta z_t,\; \|\Delta\|\leq1,
\end{align}
where
\begin{align*}
    B^w = I_{n},\;
    C_z =\begin{bmatrix}
        \rho I_{n}\\
        0
    \end{bmatrix},\;
    D_z^u &= \begin{bmatrix}
        0\\
        \rho I_{m}
    \end{bmatrix}. 
\end{align*}
The closed-loop system with a state feedback controller $u_t=-Kx_t$ thus becomes 
\begin{align}\label{closedloop_equivalent}
    x_{t+1} = \left((\hat A - \hat B K)+ \Delta (C_z-D^u_zK)\right)x_t.
\end{align}
Also, the cost function \eqref{eq:cost} becomes
\begin{align} \label{eq:cost_closedloop}
    J\left(x_0\right)=\lim _{n \rightarrow \infty} \sum_{t=0}^n x_t^\top (Q+K^\top RK) x_t.
\end{align}
Now we are ready to state the central result concerning the guaranteed cost controller for the uncertain system \eqref{eq:dyanmics_uncertainty_equivalent} under state feedback controller $u_t = kx_t$.
\begin{thm}\label{thm:GCC}
State feedback controller $u_t = -Kx_t$ is a guaranteed cost controller, according to \Cref{df:GCC}, if the following LMIs hold:
\begin{align}
\begin{split}
\left[
\begin{array}{ccccc}\label{eq:LMI_thm}
-\beta\,I_{\,n+m} & 0 & 0 & C_z X - D_z^u M & 0\\
* & -I_{n_c} & 0 & C_c X - D_c^u M & 0\\
* & * & -X & \hat A X - \hat B M & \beta I_n\\
* & * & * & -X & 0\\
* & * & * & * & -\beta\,I_{\,n}
\end{array}
\right]\ &\preceq\ 0,  \\
P&\succ 0, \\
\beta &> 0.
\end{split}
\end{align}
If the LMIs \eqref{eq:LMI_thm} hold, the state feedback gain is $K=M\,X^{-1}$.
\end{thm}
\begin{proof}
    Let $c\left(x_t, u_t\right)=x_t^\top Q x_t + u_t^\top R u_t$ and $V\left(x_t\right)=x_t^\top P x_t$ be a value function. Then from Bellman's optimality principle, we obtain
   \begin{align}\label{eq:Bellman_fin}
V_t^*\left(x_t\right)=\max _{\Delta} \min _{u_t} c\left(x_t, u_t\right)+V_{t+1}^*\left(x_{t+1}\right).
\end{align}
In the infinite horizon case, \eqref{eq:Bellman_fin} becomes
\begin{align*}
    V^*\left(x_t\right)=\max _{\Delta} \min _{u_t} c\left(x_t, u_t\right)+V^*\left(x_{t+1}\right).
\end{align*}
If \Cref{df:GCC} holds, there exists a stabilizing controller $u_t = -Kx_t$ and a sub-optimal value function $V\left(x_t\right)=x_t^\top P x_t \geq V^*\left(x_t\right)$ such that
\begin{align*}
    V\left(x_t\right)=\max _{\Delta} c\left(x_t,-K x_t\right)+V\left(x_{t+1}\right),
\end{align*}
which implies:
\begin{align} \label{eq:cost_optimality}
     V\left(x_t\right) \geq c\left(x_t,-K x_t\right)+V\left(x_{t+1}\right),\; \text{for all admissible }\Delta.
\end{align}
 Substituting $c(x_t,-Kx_t) = x_t^\top Qx_t +(-Kx_t)^\top R (-Kx_t) = x_t^\top(C_c - D_c^uK)^\top(C_c - D_c^uK)x_t$ in \eqref{eq:cost_optimality} yields
\begin{align}\label{eq:cost_optimality_expanded}
       x_t^\top(C_c - D_c^uK)^\top(C_c - D_c^uK)x_t+V\left(x_{t+1}\right) -V\left(x_t\right) \leq 0,
\end{align}
for all admissible $\Delta$.
Since $\|\Delta\|\leq 1$, we get that $w_t^\top w_t - z_t^\top z_t \leq 0$. Substituting $w_t$ and $z_t$ yields
\begin{align}\label{eq:dist_atten}
    x_t^\top \bar C_z ^\top \Delta^\top \Delta \bar C_z
 x_t - x_t^\top \bar C_z^\top \bar C_z x_t \leq 0,
 \end{align}
where $\bar C_z = \rho \begin{bmatrix}
    I_{n}\\
    -K
\end{bmatrix}$.\\
Substituting $V(x_t)$ and $V(x_{t+1})$ in \eqref{eq:cost_optimality_expanded} and applying the S-procedure \cite{Boyd1994LinearMattrix} to \eqref{eq:dist_atten} and \eqref{eq:cost_optimality_expanded}, we get: 
\begin{align} \label{eq:LMI_blocks}
    \begin{bmatrix}
        \left(\hat A-\hat BK\right)^\top P(\bullet)-P  & \left(\hat A-\hat BK\right)^\top P  \\
        P\left(\hat A-\hat BK\right) & P
    \end{bmatrix}
    &+
    \begin{bmatrix}
        (C_c - D_c^uK)^\top(\bullet) & 0 \\
        0 & 0
    \end{bmatrix} \nonumber\\
    &-\left[
    \begin{array}{cc}
       \alpha \bar C_z^\top \bar C_z  &  0\\
       0  & -\alpha
    \end{array}
    \right]\leq0,
\end{align}
where $\alpha>0$.\\
 Applying Schur complement in \eqref{eq:LMI_blocks} to the terms dependent on $P$, to the cost terms, and to the terms dependent on $\alpha$ yields:
\begin{align} \label{eq:LMI_preConjugate}
    \left[\begin{array}{ccccc}
-\alpha^{-1}I_{n+m} & 0 & 0 & \bar{C}_z & 0 \\
* & -I_{n+m} & 0 & C_c - D_c^uK & 0 \\
* & * & -P^{-1} & A-BK & I_{n} \\
* & * & * & -P & 0 \\
* & * & * & * & -\alpha I_{n}
\end{array}\right] \preceq 0.
\end{align}
 Now, we apply the congruence transformation in \eqref{eq:LMI_preConjugate} with $T= \operatorname{diag}\left(I_{n+m}, I_{n_c}, I_{n}, P^{-1}, \alpha^{-1}I_n \right)$, and perform the substitution $X=P^{-1}, \beta=\alpha^{-1}$, we obtain:
\begin{align} \label{eq:LMI_postConjugate}
    \left[\begin{array}{ccccc}
-\beta I_{n+m} & 0 & 0 & \bar{C}_zX & 0 \\
* & -I_{n+m} & 0 & C_cX - D_c^uKX & 0 \\
* & * & -X & \hat AX-\hat BKX & \beta I_{n} \\
* & * & * & -X & 0 \\
* & * & * & * & -\beta I_{n}
\end{array}\right] \preceq 0.
\end{align}
 Substituting $M=KX$, \eqref{eq:LMI_postConjugate} and \eqref{eq:LMI_thm} are equivalent. This concludes the proof.
\end{proof}

\section{Numerical Examples} \label{sec:Validation}
This section validates the preceding theoretical analysis through numerical simulations on two illustrative LTI systems: a DC motor with load and a mass-spring-damper system. We specifically investigate the impact of varying the quantization resolution on both the accuracy of the system identification and the performance of the resulting guaranteed cost controller.
\begin{figure*}
\centering 
\subfloat[]{\includegraphics[trim=1.9cm 7cm 2cm 7cm, clip=true, width=0.24\textwidth]{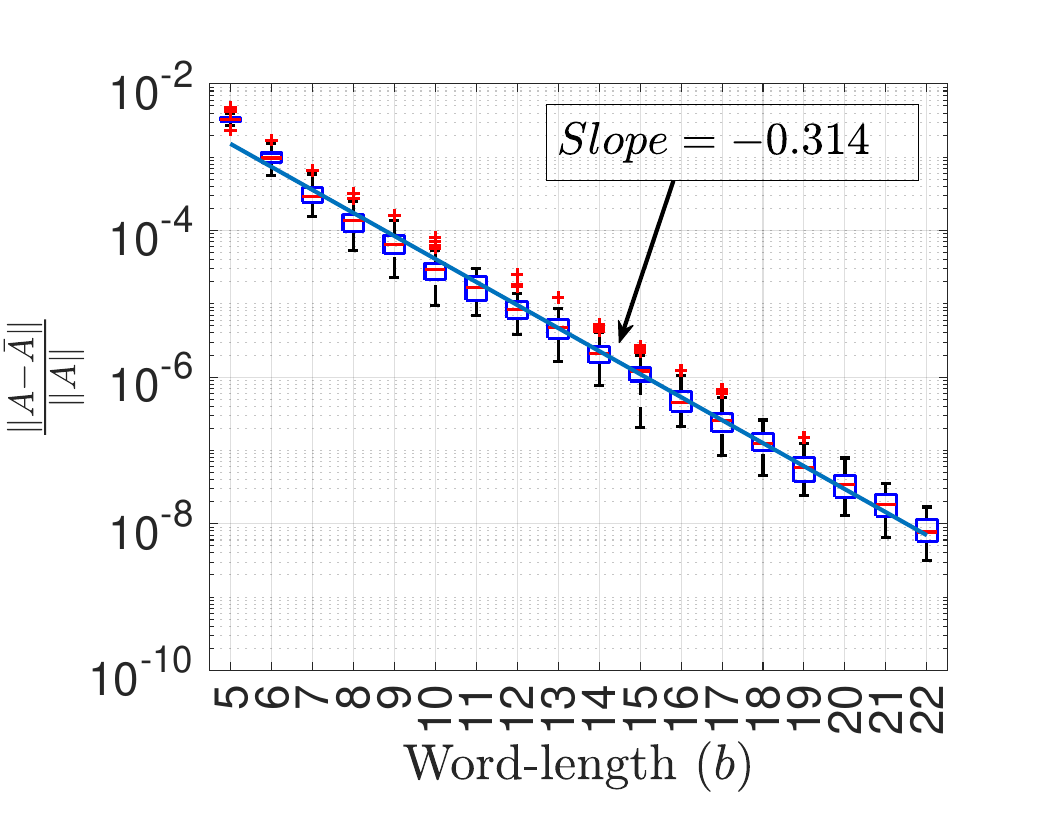}}
\subfloat[]{\includegraphics[trim=1.9cm 7cm 2cm 7cm, clip=true, width=0.24\textwidth]{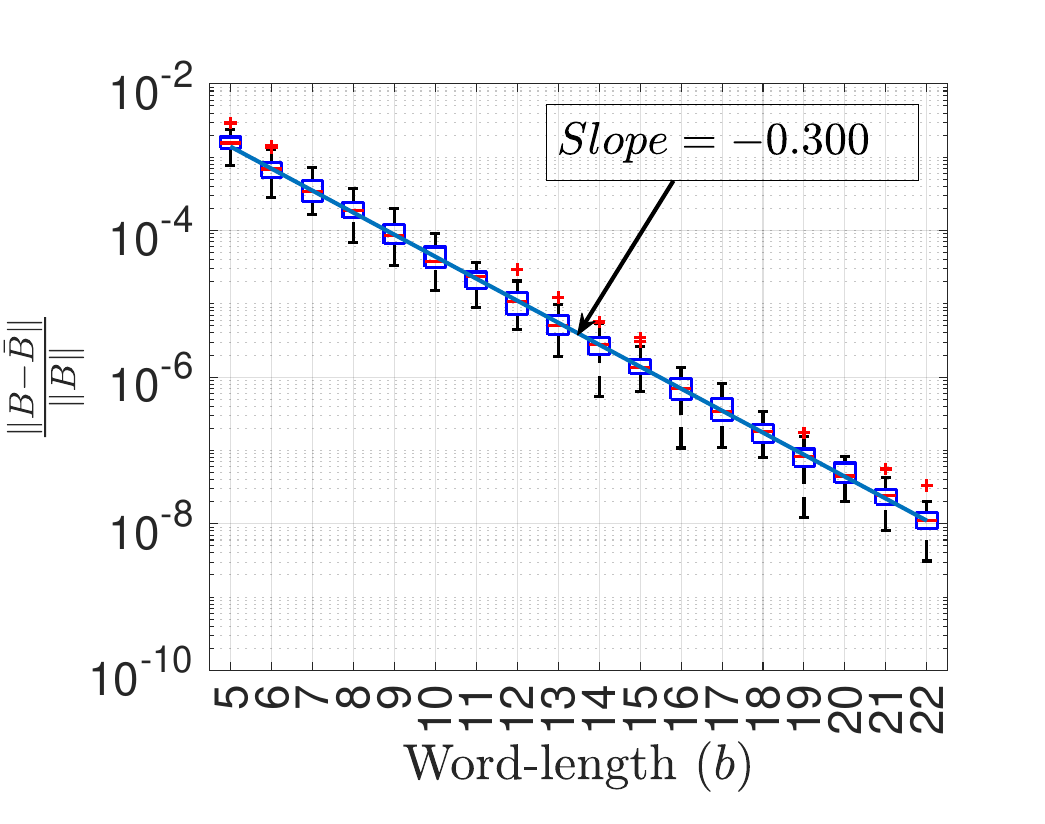}}
\subfloat[]{\includegraphics[trim=2cm 7cm 2cm 7cm, clip=true, width=0.24\textwidth]{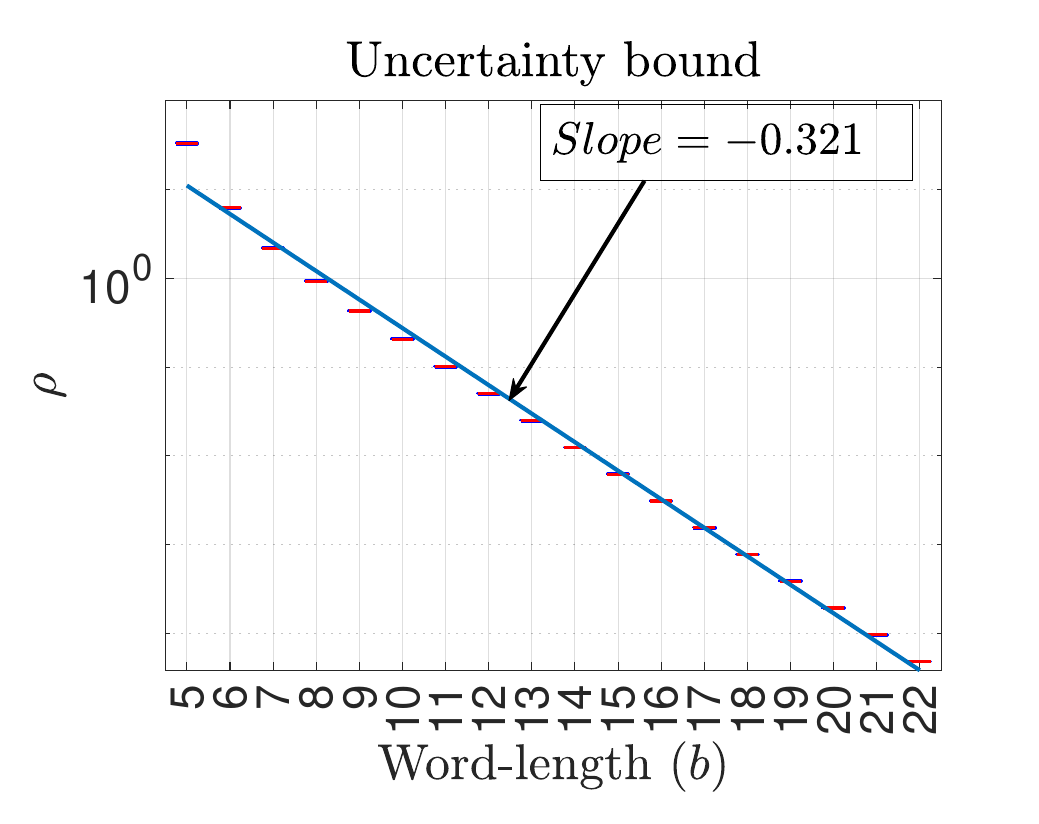}}
\subfloat[]{\includegraphics[trim=2cm 7cm 2cm 7cm, clip=true, width=0.24\textwidth]{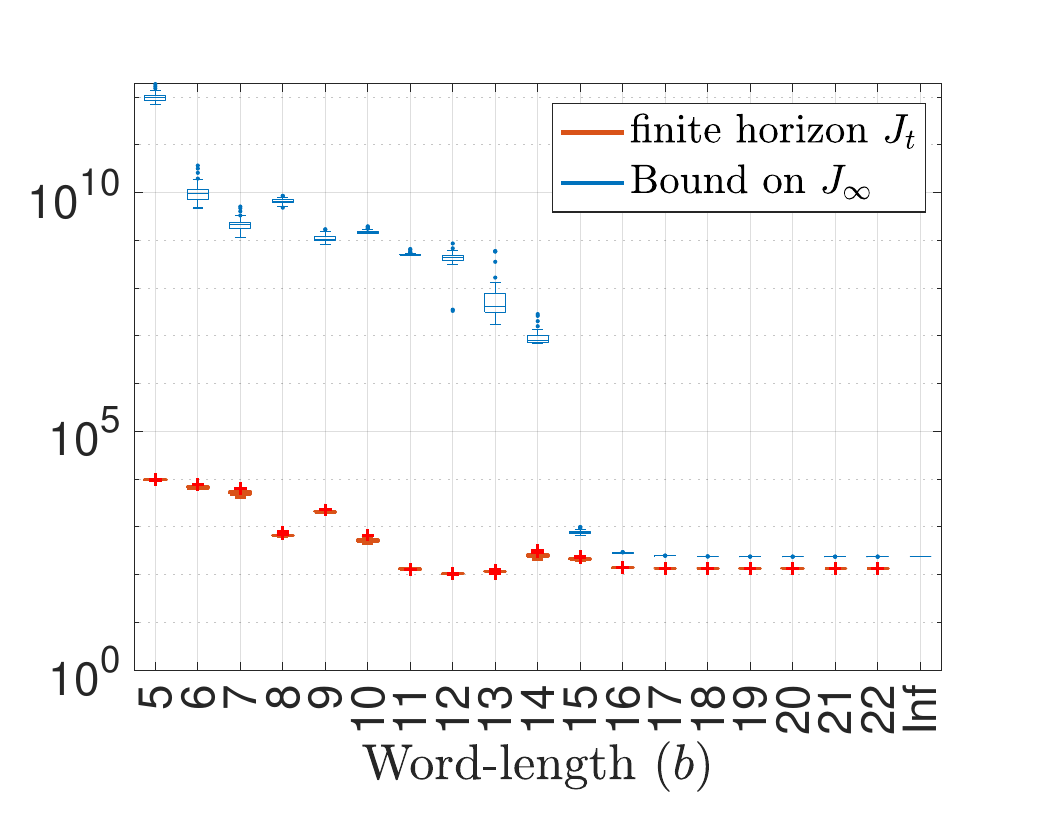}}

\subfloat[]{\includegraphics[trim=2cm 7cm 2cm 7cm, clip=true, width=0.24\textwidth]{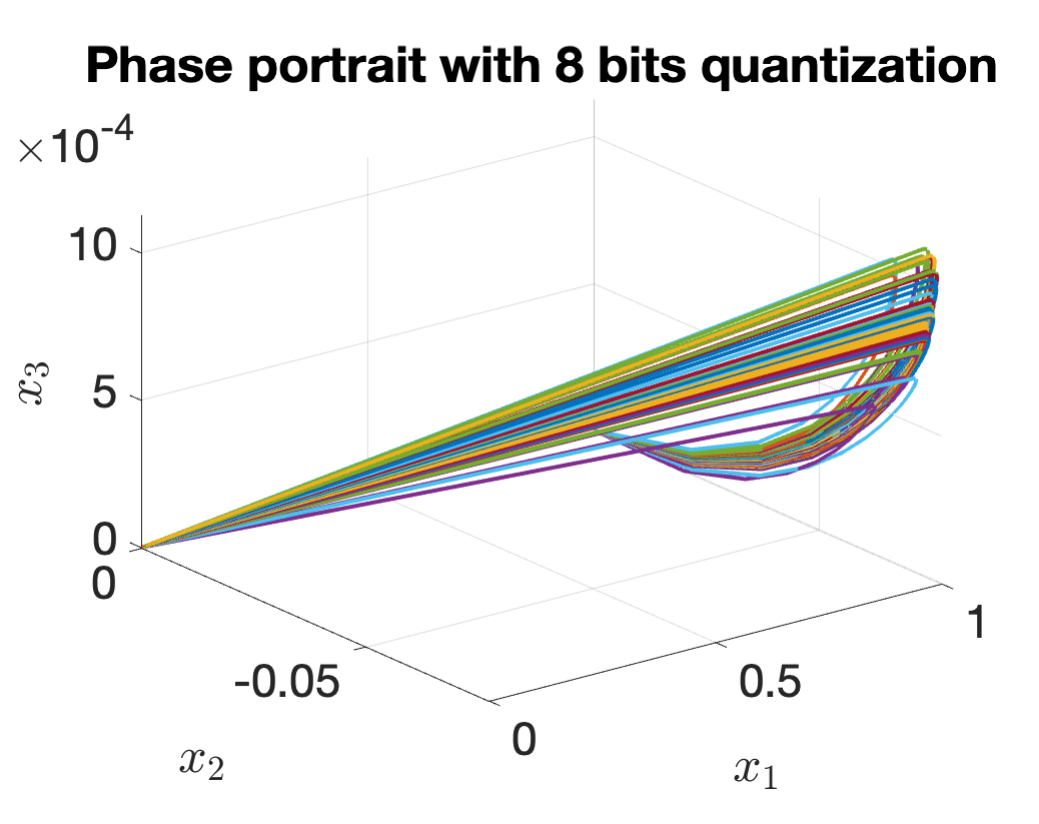}}
\subfloat[]{\includegraphics[trim=2cm 7cm 2cm 7cm, clip=true, width=0.24\textwidth]{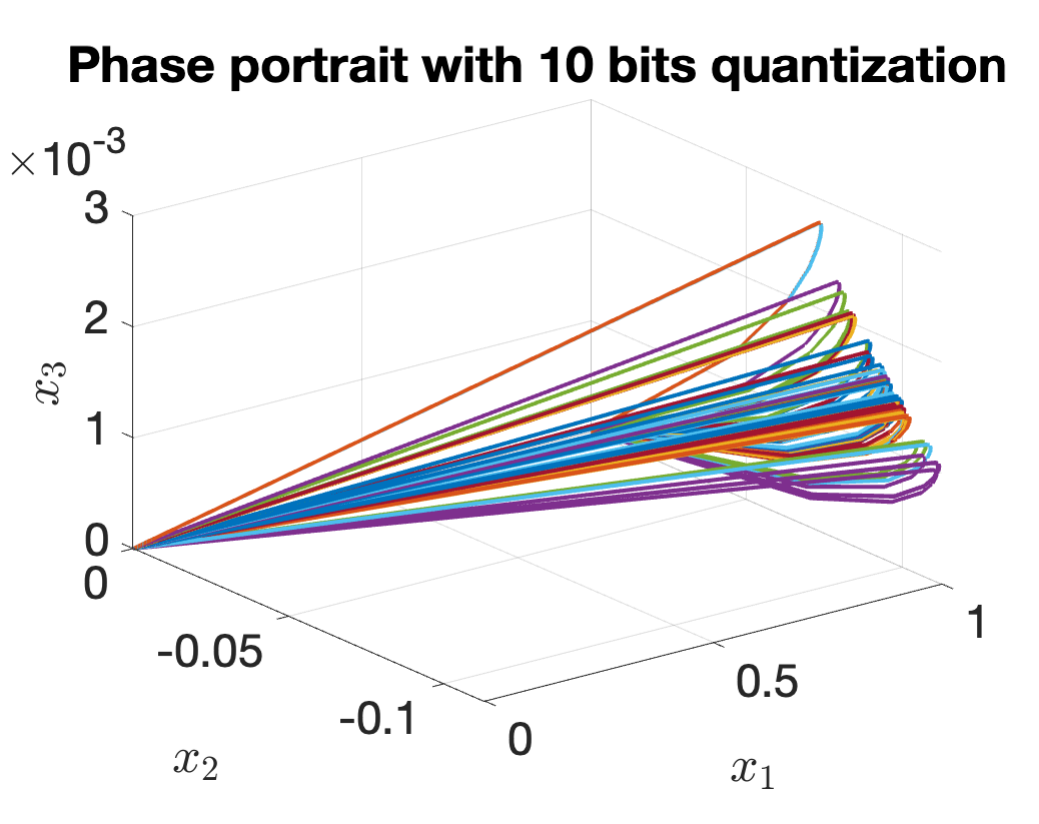}}
\subfloat[]{\includegraphics[trim=2cm 7cm 2cm 7cm, clip=true, width=0.24\textwidth]{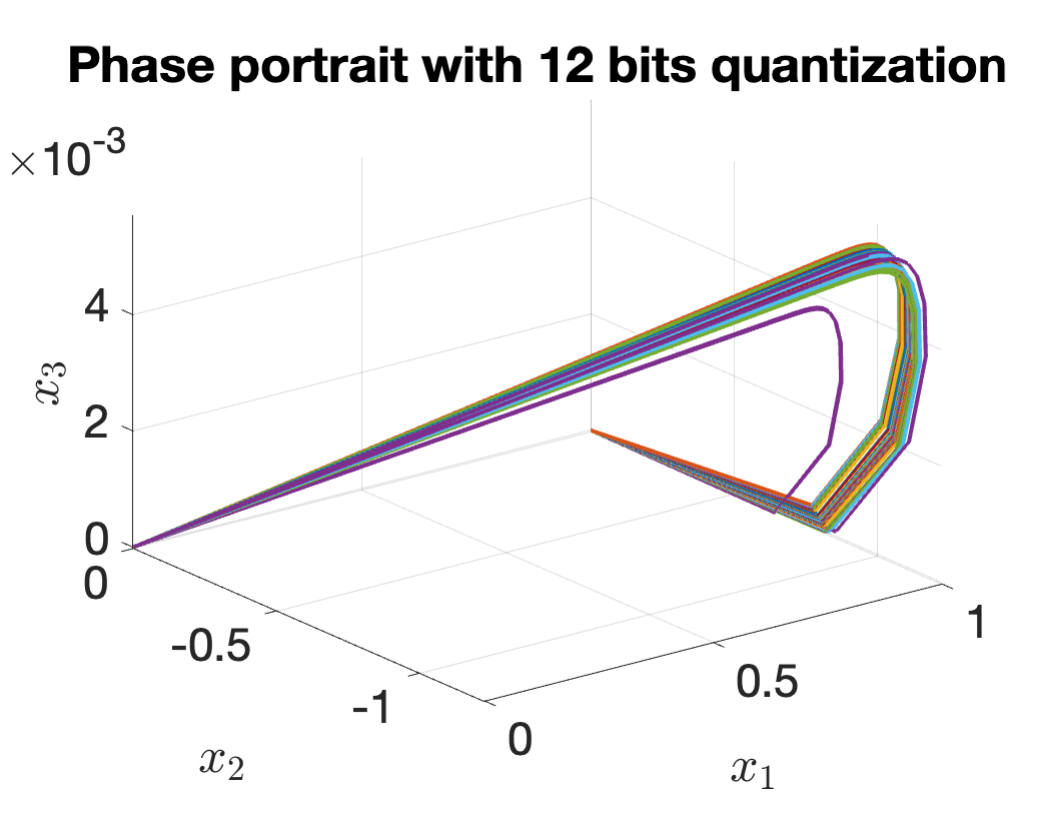}}
\subfloat[]{\includegraphics[trim=2cm 7cm 2cm 7cm, clip=true, width=0.24\textwidth]{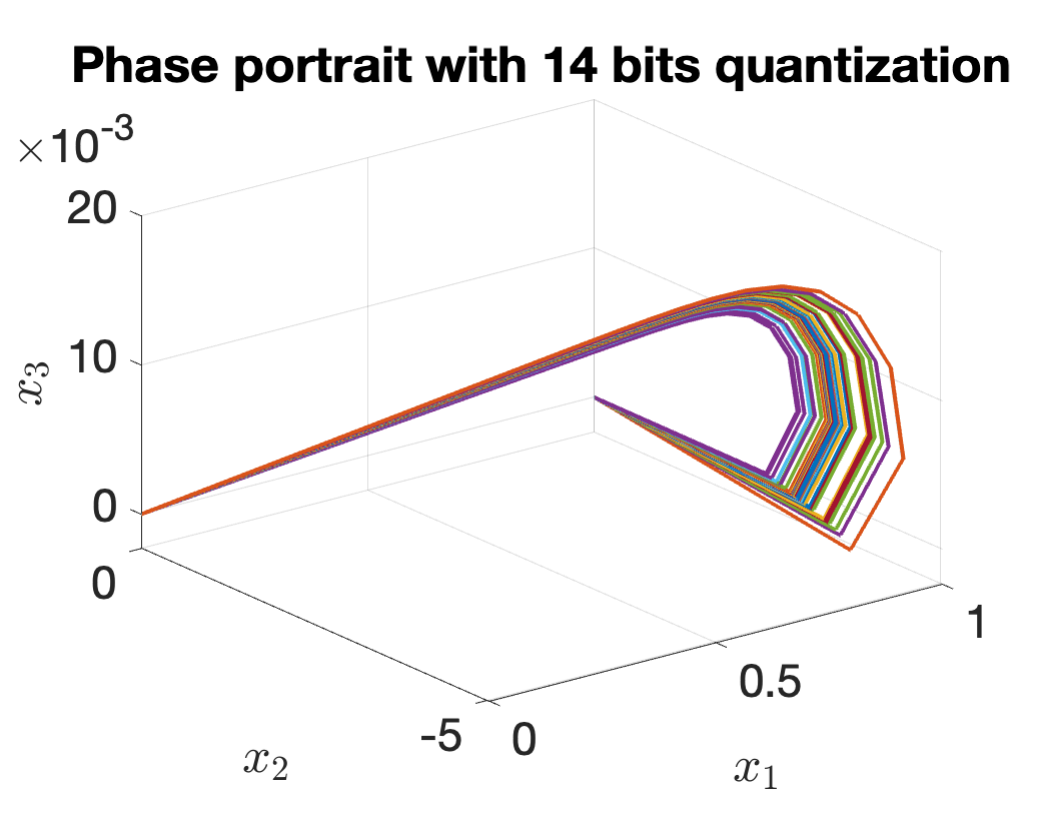}}
\caption{Error and phase-portrait profile for DC Motor with Load: (a) relative error in matrix $A$; (b) relative error in matrix $B$; (c) uncertainty norm bound $\rho$; (d) finite-horizon cost and infinite-horizon cost bound; (e)--(h) phase portrait from regulation (with models identified from data snapshots generated by 50 independent random input and initial condition realization) for world-lengths $b=8,\ 10, \ 12,$ and $ 14$, respectively.}
\label{Fig: Motor} \vspace{-0.2cm}
\end{figure*}
\subsection{DC Motor with Load}
A DC Motor with control torque on the load is considered for this example. The dynamics of the system are as follows: \vspace{-14pt}
\begin{align}\label{eq: Motor}
\begin{split}
    \dot{x}_1 = x_2,\
\dot{x}_2 = \tfrac{K_t\,x_3 - b\,x_2 + u_2}{J},\
\dot{x}_3 = \tfrac{u_1 - R\,x_3 - K_e\,x_2}{L}.
\end{split}
\end{align}
Using motor parameters $J = 0.01$, $b = 0.2$, $K_t = K_e = 0.01$, $R = 1$, and $L = 0.02$, and discretizing with a 0.01 second interval yields:
\vspace{-2pt}
\renewcommand{\arraystretch}{0.85}
\begin{align*}
\textstyle
    A&=\left[
    \begin{array}{ccc}
        1.000 & 0.0091 &  0\\
        0 & 0.8187 & 0.0071 \\
        0 & -0.0035 & 0.6065
    \end{array}
    \right],\\
    B&= \left[
    \begin{array}{cc}
        0 & ~~0.0047 \\
        0.0020 & ~~0.9063  \\
        0.3935 & -0.0020
    \end{array}
    \right].
\end{align*}
\renewcommand{\arraystretch}{1.0}
For training, the initial conditions are generated randomly with uniform distribution in the cube $[-0.1, 0.1]^3$. The first and second control inputs for each trajectory are chosen to be a uniformly distributed random signal on $[-0.2, 0]$ and $[-0.1, 0.1]$, respectively. For data collection, the system is simulated for 150 trajectories over 100 sampling periods (i.e., 1 second per trajectory). The identification phase is then repeated 50 times.  The robust persistence of excitation condition \eqref{eq:PEcondition} was met for all quantization word-lengths of 5 bits and higher. Relative 2-norm error $\tfrac{\| A - \hat{A}\|}{\|A\|}$ and $\tfrac{\| B - \hat{B}\|}{\|B\|}$ for different word-length are shown in Fig.~\ref{Fig: Motor}(a)-(b). 
Also, the uncertainty norm bounds obtained for different word-lengths are demonstrated in Fig.~\ref{Fig: Motor}(c). A guaranteed cost controller is then designed based on the identified model $[\hat{A}, \hat{B}]$ for the performance objective defined in \eqref{eq:cost}, with $Q = \operatorname{diag}([1, 1, 1])$ and $R = \operatorname{diag}([1, 1])$. We then evaluate the controller's effectiveness by simulating the regulation of the system from the initial state $x_0=[1, 0, 0]^\top$ to the origin. The finite-horizon cost function values and infinite-horizon cost bounds achieved for different word lengths are demonstrated in Fig.~\ref{Fig: Motor}(d). Figs.~\ref{Fig: Motor}(e)--(h) show the phase-portrait of controlled state-trajectories for different word-lengths. The numerical results show a linear relationship between the logarithm of the identification errors and the quantization word-length $b$. We observe that the logarithmic errors in the estimated matrices $\hat{A}$ and $\hat{B}$, along with the uncertainty norm bound $\rho$, decrease with slopes of approximately -0.314, -0.300, and -0.321, respectively. This empirical finding aligns with the theoretical analysis in \Cref{thm:bounded_uncertainty}, which predicts that the error should scale as $O(\varepsilon) \propto 2^{-b}$, implying a theoretical slope of $-\log_{10}(2) \approx -0.301$.
 Also, as analytically predicted, higher word-length $b$ (lower quantization resolution $\epsilon$) reduces guaranteed cost and improves trajectory convergence, evident from the close alignment of all 50 Monte-Carlo trajectories for $b=12$, and $14$.

\begin{figure*}
\centering 
\subfloat[]{\includegraphics[trim=1.9cm 7cm 2cm 7cm, clip=true, width=0.24\textwidth]{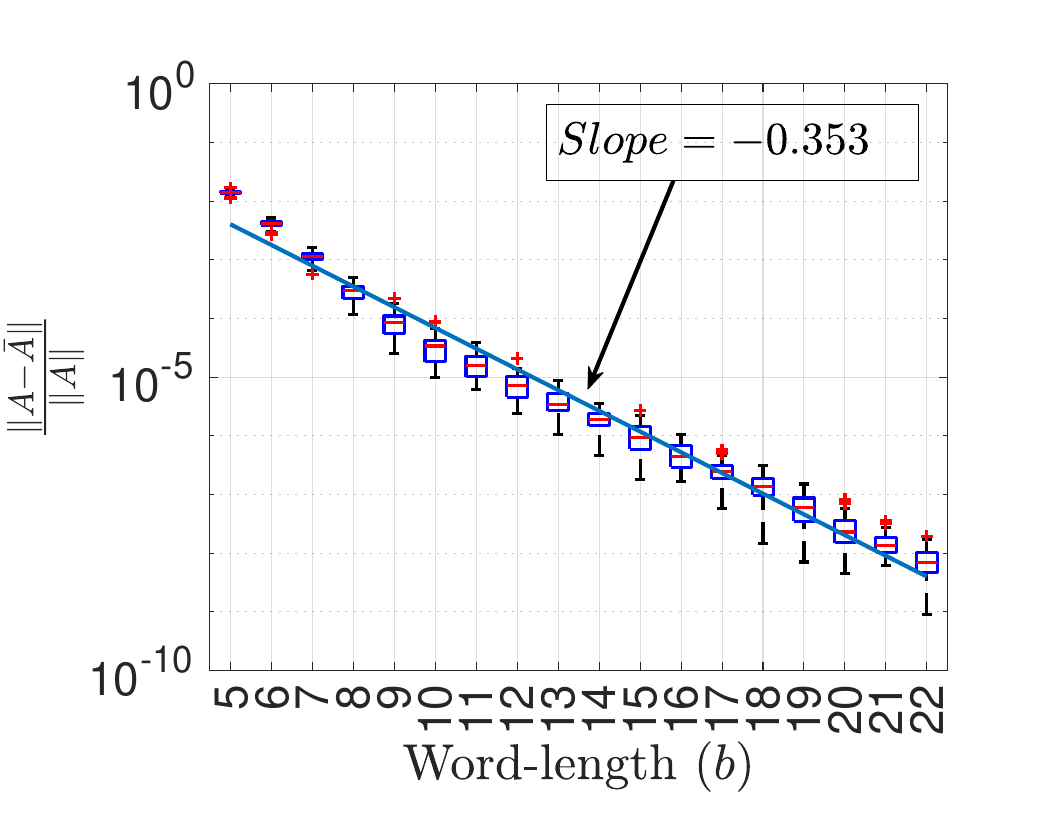}}
\subfloat[]{\includegraphics[trim=1.9cm 7cm 2cm 7cm, clip=true, width=0.24\textwidth]{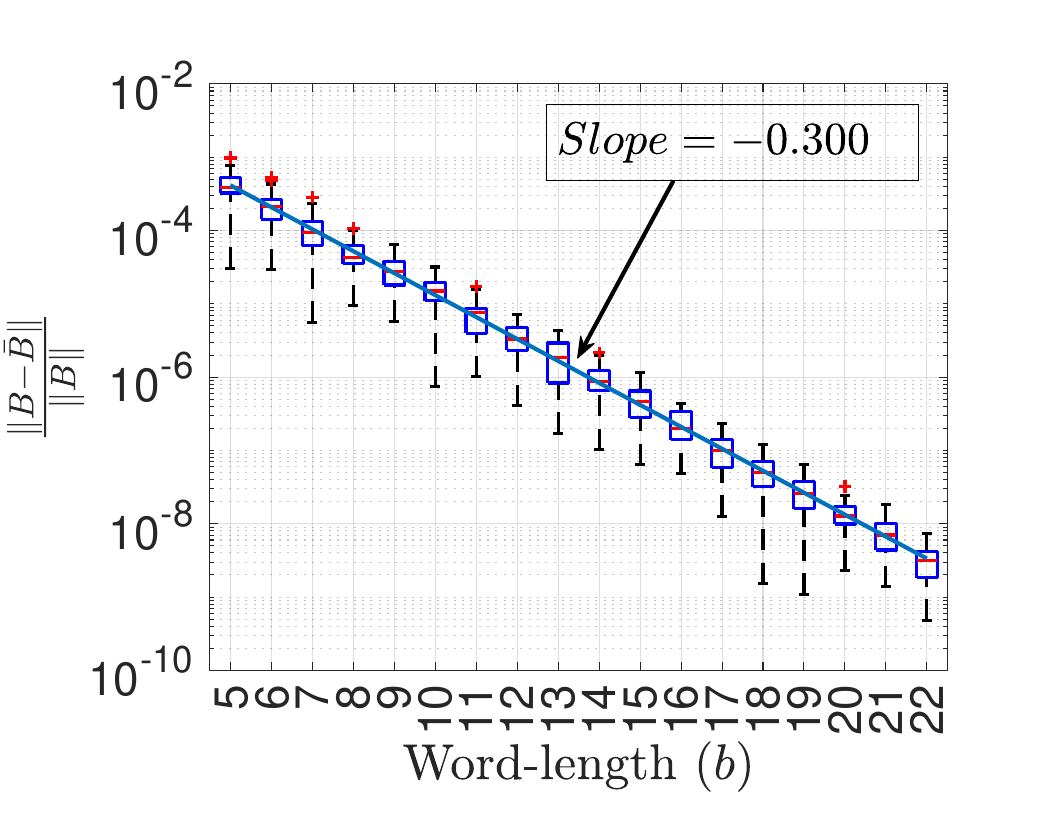}}
\subfloat[]{\includegraphics[trim=2cm 7cm 2cm 7cm, clip=true, width=0.24\textwidth]{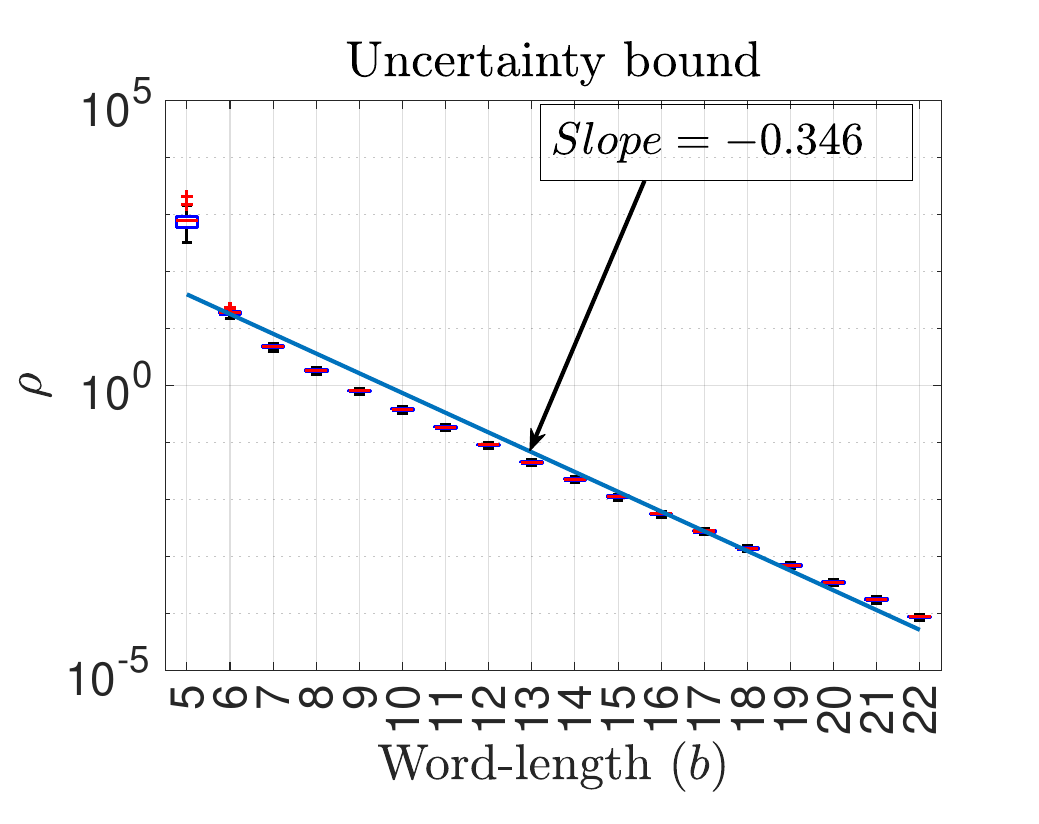}}
\subfloat[]{\includegraphics[trim=2cm 7cm 2cm 7cm, clip=true, width=0.24\textwidth]{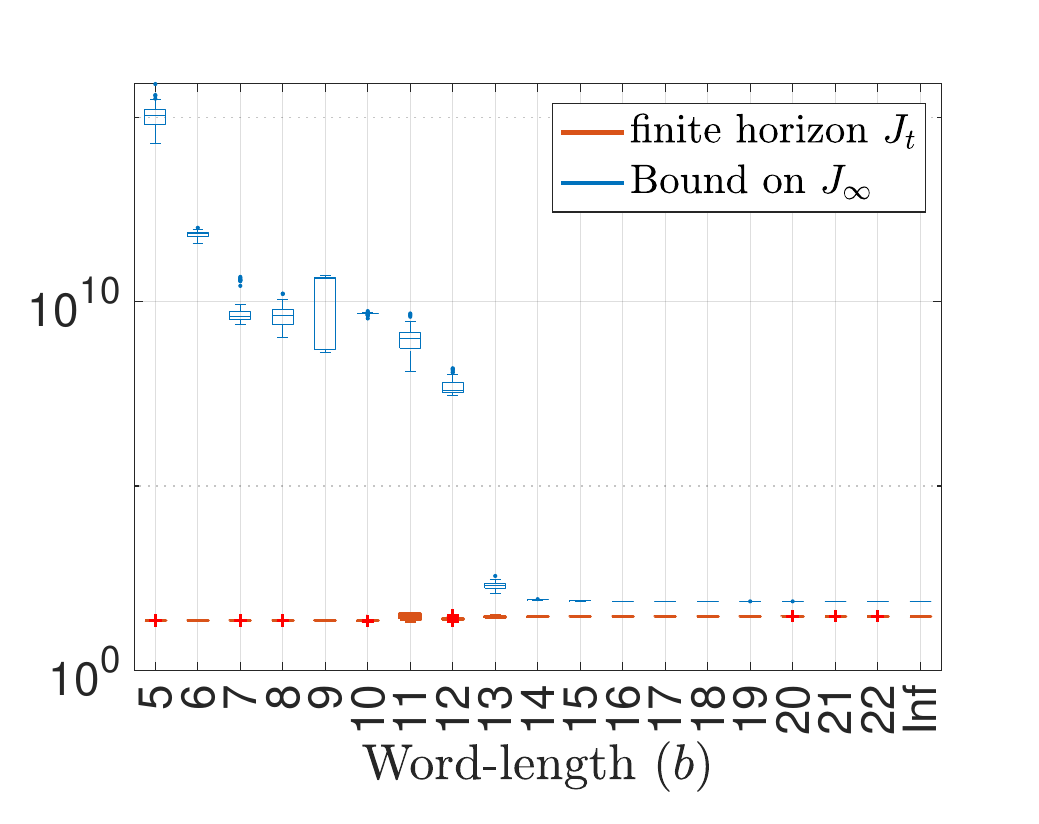}}\\ \vspace{-3pt}
\subfloat[]{\includegraphics[trim=2cm 7cm 2cm 7cm, clip=true, width=0.24\textwidth]{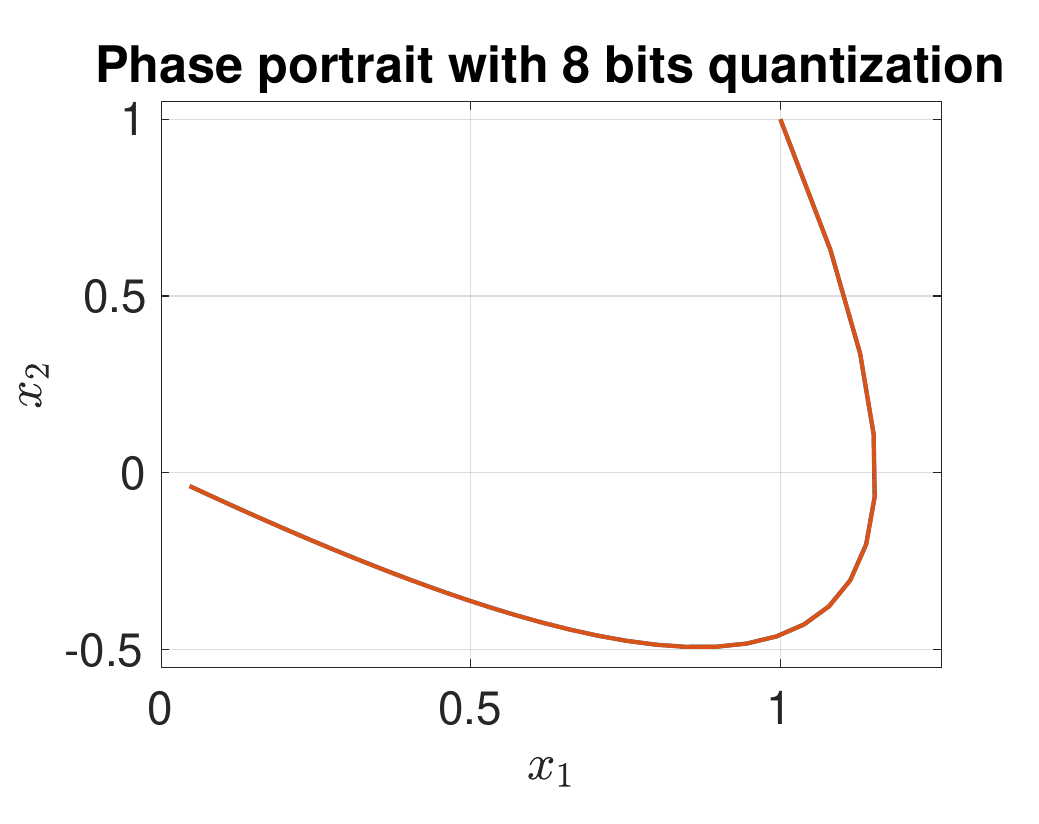}}
\subfloat[]{\includegraphics[trim=2cm 7cm 2cm 7cm, clip=true, width=0.24\textwidth]{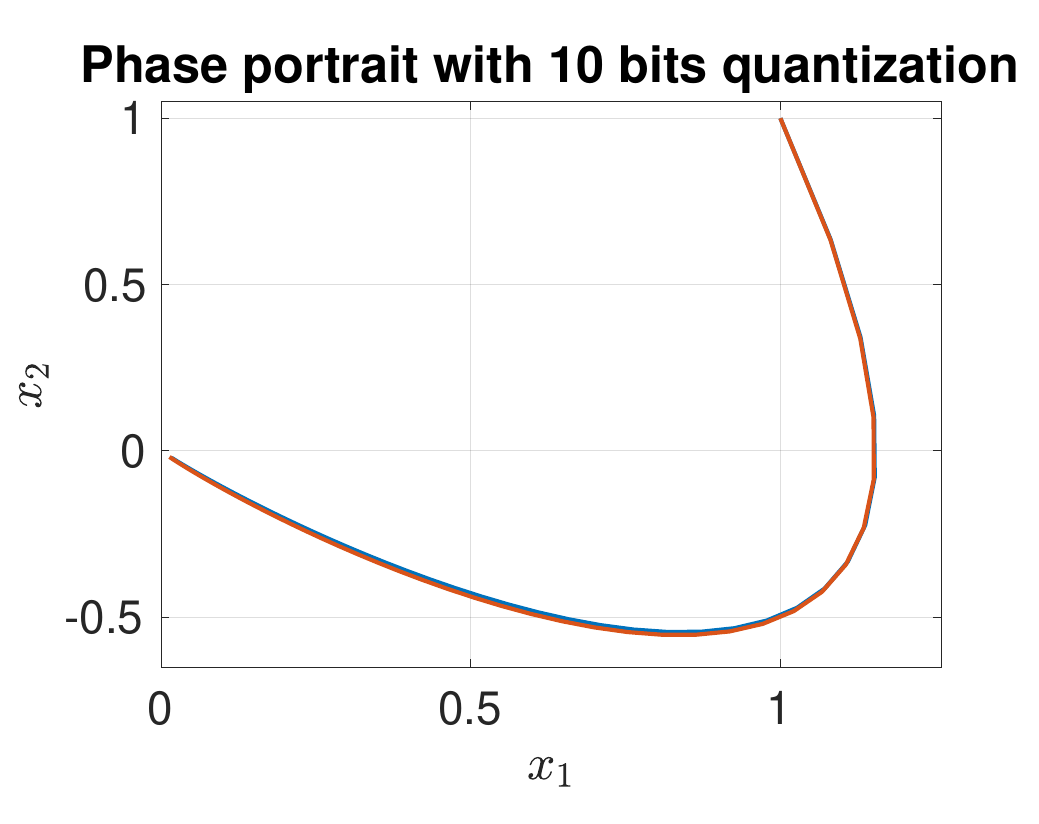}}
\subfloat[]{\includegraphics[trim=2cm 7cm 2cm 7cm, clip=true, width=0.24\textwidth]{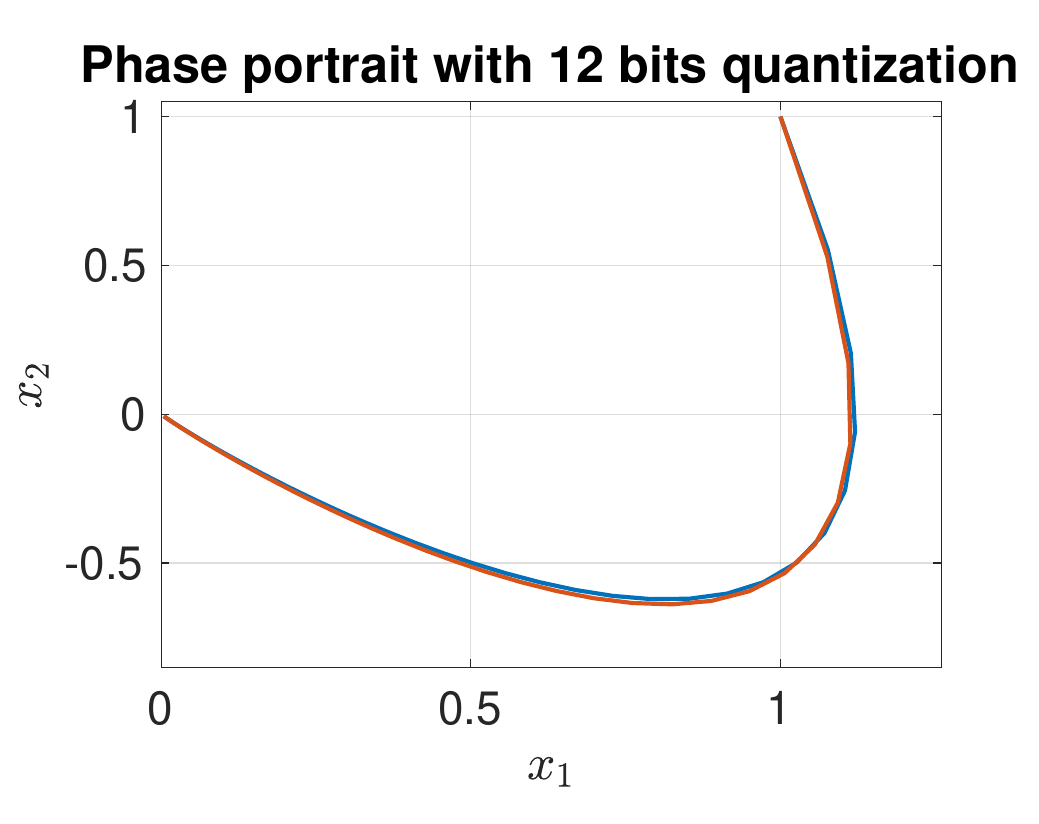}}
\subfloat[]{\includegraphics[trim=2cm 7cm 2cm 7cm, clip=true, width=0.24\textwidth]{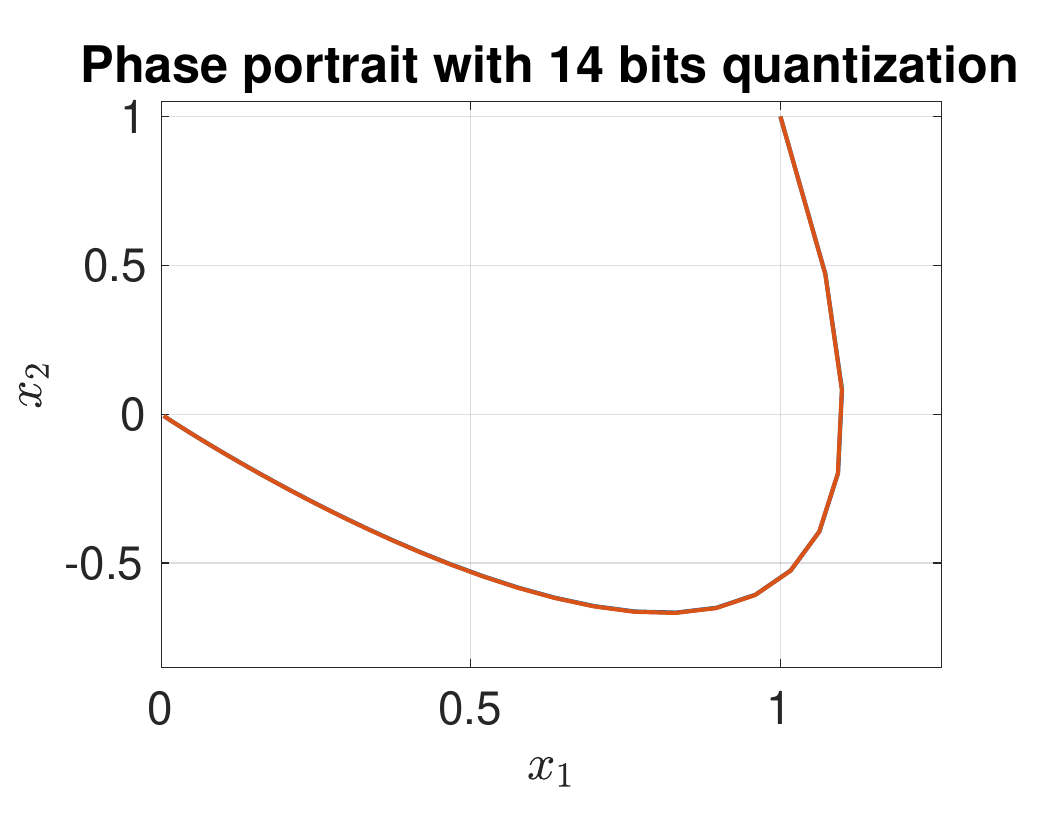}}
\caption{Error and phase-portrait profile for mass-spring-damper: (a) relative error in matrix $A$; (b) relative error in matrix $B$; (c) uncertainty norm bound $\rho$; (d) finite-horizon cost and infinite-horizon cost bound; (e)--(h) phase portrait from regulation (with models identified from data snapshots generated by 50 independent random input and initial condition realization) for world-lengths $b=8,\ 10,\ 12,$ and $ 14$, respectively.}
\label{Fig: MSD} \vspace{-0.2cm}
\end{figure*}

\subsection{Mass Spring Damper}
The second example is a mass-spring-damper system. The dynamics of the system are as follows: 
\begin{align}\label{eq: massspring}
\begin{split}
    \dot{x}_1 = x_2,\
\dot{x}_2 = \frac{-kx_1-cx_2+u}{m}
\end{split}
\end{align}
Using physical parameters $m=1,\, k=1.5,\, c=2.5$ and discretizing with a 0.1 second interval yields:
\begin{align*}
\textstyle
    A=\left[
    \begin{array}{cc}
         0.9931  &  0.0883  \\
         -0.1324 &  0.7724
    \end{array}
    \right],\;
    B= \left[
    \begin{array}{c}
        0.0046\\
        0.0883
    \end{array}
    \right].
\end{align*}

For training, the initial conditions are generated randomly with uniform distribution in the square $[-0.2, 0.2]^2$. The input for each trajectory is chosen to be a uniformly distributed random signal on  $[-0.1, 0.1]$. The rest of the training settings remain the same as that of the DC motor with load. Figs.~\ref{Fig: MSD}(a), (b), and (c) show similar trends for errors in linear predictor matrices $A$ and $B$ and uncertainty bound $\rho$. A guaranteed cost controller is then designed based on the identified model $[\hat{A}, \hat{B}]$ for the performance objective defined in \eqref{eq:cost}, with $Q = \operatorname{diag}([1\ 1])$, $R = 1$. We then evaluate the controller's effectiveness by simulating the regulation of the system from the initial state $x_0=[1, 1]^\top$ to the origin. The finite-horizon cost function values and infinite-horizon cost bounds achieved for different word lengths are demonstrated in Fig.~\ref{Fig: MSD}(d). Figs.~\ref{Fig: MSD}(e)--(h) show the phase-portrait of controlled state trajectories for different word-lengths. We see a similar trend in identification error, uncertainty norm bound, and guaranteed cost here as well.

\section{Conclusions} \label{sec:conclusions}
In this paper, we investigate the effect of input and state data quantization on LTI system identification and subsequent design of a guaranteed cost robust controller. We theoretically establish a fundamental norm bound on the identification error between true and identified system matrices. This error bound depends \emph{solely on the available quantized data} and quantization resolution and assumes no knowledge of true system matrices. Furthermore, we develop a linear matrix inequality (LMI) based guaranteed cost stabilizing controller that is robust under the aforementioned model uncertainty pertaining to the identification error. Our analysis is validated via repeated experiments on multiple problems.

\bibliographystyle{ieeetr}
\bibliography{ref1, references,revisionref}

\begin{thebibliography}{10}

\bibitem{zarei2025MAS}
F.~Zarei and B.~Shafai, ``Consensus of multi-agent systems under disturbances
  using a proportional-integral fading observer,'' in {\em Integrated Systems:
  AI-Augmented Engineering}, Springer.

\bibitem{Vaziri2025CNN}
A.~Vaziri and H.~Fang, ``Optimal inferential control of convolutional neural
  networks,'' in {\em 2025 American Control Conference (ACC)}, pp.~2603--2610,
  2025.

\bibitem{hedesh2024NN}
H.~Montazeri~Hedesh and M.~Siami, ``Ensuring both positivity and stability
  using sector-bounded nonlinearity for systems with neural network
  controllers,'' {\em IEEE Control Systems Letters}, vol.~8, pp.~1685--1690,
  2024.

\bibitem{alavi2025safety}
A.~Alavi, A.~Nadali, M.~Zamani, and S.~Jafarpour, ``Neural barrier certificates
  for monotone systems,'' {\em IEEE Control Systems Letters}, vol.~9,
  pp.~1496--1501, 2025.

\bibitem{Cleary2020}
A.~Cleary, K.~Yoo, P.~Samuel, S.~George, F.~Sun, and S.~A. Israel, ``Machine
  learning on small uavs,'' in {\em 2020 IEEE Applied Imagery Pattern
  Recognition Workshop (AIPR)}, pp.~1--5, 2020.

\bibitem{Li2023cloud}
N.~Li, K.~Zhang, Z.~Li, V.~Srivastava, and X.~Yin, ``Cloud-assisted nonlinear
  model predictive control for finite-duration tasks,'' {\em IEEE Transactions
  on Automatic Control}, vol.~68, no.~9, pp.~5287--5300, 2023.

\bibitem{Schweickhardt2009Linearization}
T.~Schweickhardt and F.~Allgower, ``On system gains, nonlinearity measures, and
  linear models for nonlinear systems,'' {\em IEEE Transactions on Automatic
  Control}, vol.~54, no.~1, pp.~62--78, 2009.

\bibitem{korda2018}
M.~Korda and I.~Mezi\'{c}, ``{Optimal construction of Koopman eigenfunctions
  for prediction and control},'' {\em ArXiv e-prints}, 2018.

\bibitem{hedesh2025LMI}
H.~M. Hedesh, M.~K. Wafi, and M.~Siami, ``Local stability and region of
  attraction analysis for neural network feedback systems under positivity
  constraints,'' 2025.

\bibitem{Ganji2024LMI}
M.~Ganji and M.~Pourgholi, ``An lmi-based robust fuzzy controller for blood
  glucose regulation in type 1 diabetes,'' in {\em 2024 32nd International
  Conference on Electrical Engineering (ICEE)}, pp.~1--7, 2024.

\bibitem{Zhang2013NCSConstraints}
L.~Zhang, H.~Gao, and O.~Kaynak, ``Network-induced constraints in networked
  control systems—a survey,'' {\em IEEE Transactions on Industrial
  Informatics}, vol.~9, no.~1, pp.~403--416, 2013.

\bibitem{Danaee2018QuantizationEnergy}
A.~Danaee, R.~C. de~Lamare, and V.~H. Nascimento, ``Energy-efficient
  distributed learning with coarsely quantized signals,'' {\em IEEE Signal
  Processing Letters}, vol.~28, pp.~329--333, 2021.

\bibitem{ljung1998system}
L.~Ljung, {\em System Identification: Theory for the User}.
\newblock Pearson Education, 1998.

\bibitem{Arbabi2017}
H.~Arbabi and I.~Mezi\'{c}, ``Ergodic theory, dynamic mode decomposition, and
  computation of spectral properties of the koopman operator,'' {\em SIAM
  Journal on Applied Dynamical Systems}, vol.~16, no.~4, pp.~2096--2126, 2017.

\bibitem{Ziemann2023}
I.~Ziemann, A.~Tsiamis, B.~Lee, Y.~Jedra, N.~Matni, and G.~J. Pappas, ``A
  tutorial on the non-asymptotic theory of system identification,'' in {\em
  2023 62nd IEEE Conference on Decision and Control (CDC)}, pp.~8921--8939,
  2023.

\bibitem{maity2021optimal}
D.~Maity and P.~Tsiotras, ``Optimal controller synthesis and dynamic quantizer
  switching for linear-quadratic-{G}aussian systems,'' {\em IEEE Transactions
  on Automatic Control}, vol.~67, no.~1, pp.~382--389, 2021.

\bibitem{maity2023optimal}
D.~Maity and P.~Tsiotras, ``Optimal quantizer scheduling and controller
  synthesis for partially observable linear systems,'' {\em SIAM Journal on
  Control and Optimization}, vol.~61, no.~4, pp.~2682--2707, 2023.

\bibitem{maity2024effect}
D.~Maity, D.~Goswami, and S.~Narayanan, ``On the effect of quantization on
  dynamic mode decomposition,'' in {\em 2024 IEEE 63rd Conference on Decision
  and Control (CDC)}, pp.~8778--8785, 2024.

\bibitem{maity2024EDMD}
D.~Maity and D.~Goswami, ``On the effect of quantization on extended dynamic
  mode decomposition,'' in {\em American Control Conference}, pp.~3176--3182,
  2025.

\bibitem{ataei2025koopman}
S.~Ataei, D.~Maity, and D.~Goswami, ``Koopman meets limited bandwidth: Effect
  of quantization on data-driven linear prediction and control of nonlinear
  systems,'' {\em arXiv preprint arXiv:2501.07714}, 2025.

\bibitem{ataei2025qsid}
S.~Ataei, D.~Maity, and D.~Goswami, ``{QSID-MPC: Model predictive control with
  system identification from quantized data},'' {\em IEEE Control Systems
  Letters}, vol.~9, pp.~1880--1885, 2025.

\bibitem{schmid2010}
P.~J. Schmid, ``Dynamic mode decomposition of numerical and experimental
  data,'' {\em Journal of Fluid Mechanics}, vol.~656, pp.~5--28, 2010.

\bibitem{fu2024tutorial}
M.~Fu, ``A tutorial on quantized feedback control,'' {\em IEEE/CAA Journal of
  Automatica Sinica}, vol.~11, no.~1, pp.~5--17, 2024.

\bibitem{Boyd1994LinearMattrix}
S.~Boyd, L.~El~Ghaoui, E.~Feron, and V.~Balakrishnan, {\em Linear Matrix
  Inequalities in System and Control Theory}.
\newblock Philadelphia, PA, USA: SIAM, 1994.

\end{thebibliography}


\end{document}